\documentclass[aps,reprint,physrev,superscriptaddress]{revtex4-2}
\usepackage{blindtext}
\usepackage{amsmath}
\usepackage{amsfonts}
\usepackage{amssymb}
\usepackage{graphicx}
\usepackage{bm}
\usepackage{hyperref}
\usepackage{siunitx}
\usepackage{upgreek}

\usepackage{xcolor}

\begin{document}
\title{Mechanisms enabling efficient fountain-flow supracellular migration}

\author{Jordi Font-Reverter}
\affiliation{Universitat Polit\`ecnica de Catalunya-BarcelonaTech, 08034 Barcelona, Spain.}
\author{Alejandro Torres-S\'anchez}
\thanks{Present address: European Molecular Biology Laboratory (EMBL-Barcelona), 08003 Barcelona, Spain}
\affiliation{Universitat Polit\`ecnica de Catalunya-BarcelonaTech, 08034 Barcelona, Spain.}
\thanks{European Molecular Biology Laboratory (EMBL-Barcelona), 08003 Barcelona, Spain}
\author{Guillermo Vilanova}
\affiliation{Universitat Polit\`ecnica de Catalunya-BarcelonaTech, 08034 Barcelona, Spain.}
\author{Marino Arroyo}
\email{marino.arroyo@upc.edu}
\affiliation{Universitat Polit\`ecnica de Catalunya-BarcelonaTech, 08034 Barcelona, Spain.}
 \affiliation{Institute for Bioengineering of Catalonia (IBEC), The Barcelona Institute of Science and Technology (BIST), 08028 Barcelona, Spain.}
\affiliation{Centre Internacional de Mètodes Numèrics en Enginyeria (CIMNE), 08034 Barcelona, Spain.}

\begin{abstract}
In a prototypical mode of single-cell migration, retrograde cytoskeletal flow is mechanically coupled to the environment, propels the cell, and is sustained by an anterograde cytosolic flow of disassembled cytoskeletal components. Supracellular collectives also develop fountain-flows to migrate, but the opposing cellular streams interact with the environment producing conflicting forces. To understand the biophysical constraints of fountain-flow supracellular migration, we develop an active gel model of a cell cluster driven by a polarized peripheral contractile cable. While the model develops fountain-flows and directed migration, efficiency and cluster velocity are extremely small compared to observations. We find that patterned friction or cluster-polarized single-cell directed migration, both suggested by contact inhibition of locomotion, rescue robust and efficient supracellular migration.
\end{abstract}

\maketitle

\section{Introduction}

\begin{figure}
	\centering
	\includegraphics[width=.99\linewidth]{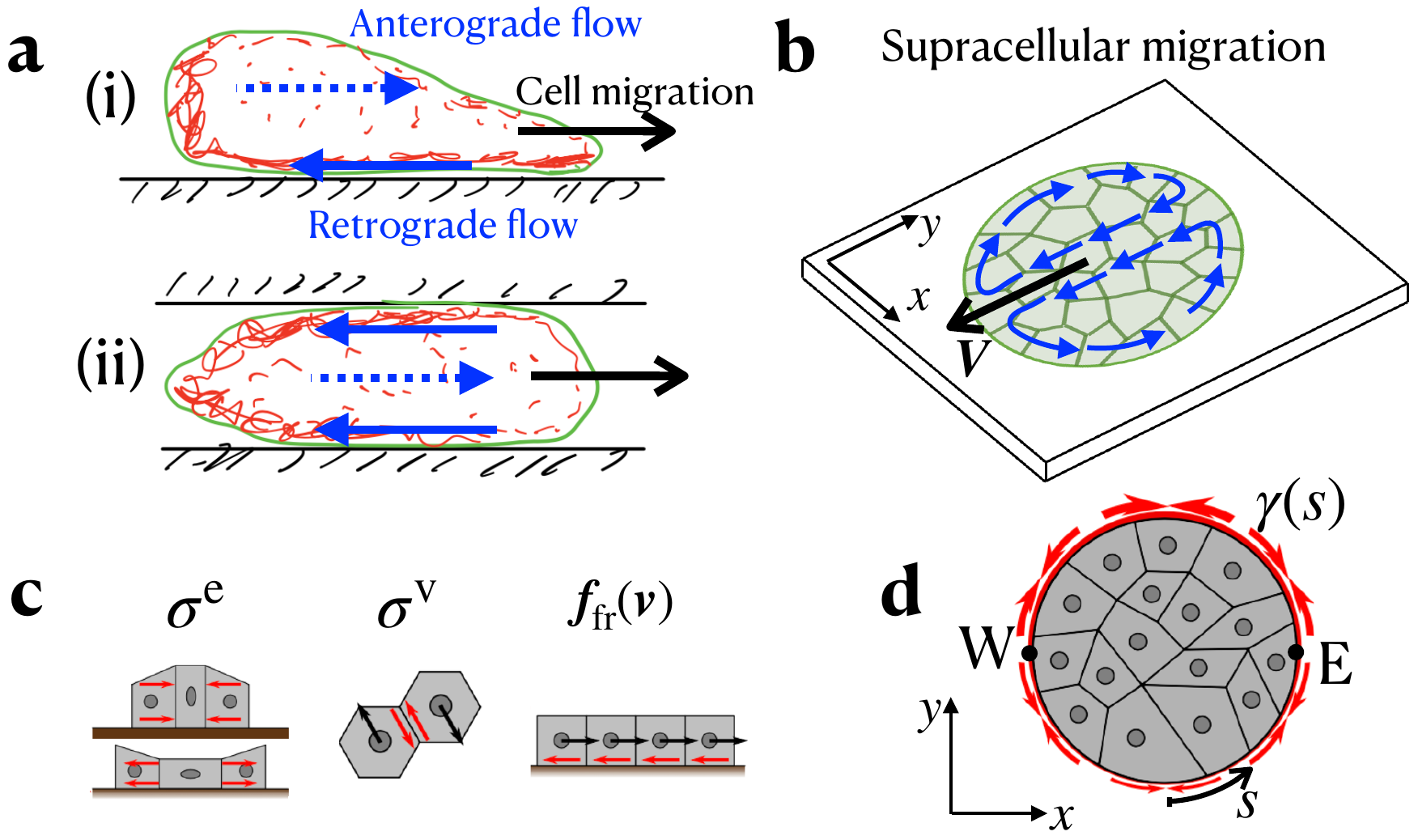}
	\caption{{\bf Fountain flow migration and main ingredients of the theoretical model.} (a) During fountain flow migration of single cells either (i) specifically adhered to a substrate or (ii) nonadherent and confined, retrograde flow of cytoskeletal materials (solid blue arrow) couples with a substrate and propels cell forward, whereas anterograde flow takes place within the cytosol (dashed blue arrow). (b) During supracellular fountain-flow migration, both retrograde and anterograde flows are mechanically coupled to the substrate. (c) Sketch of the mechanical ingredients within a cell collective including elastic stresses, viscous stresses and friction. (d) Polarized contractility $\gamma$ of a peripheral supracellular cable. }
	\label{fig1}
\end{figure}

Fountain flow migration is a prototypical mode of cellular locomotion, by which the retrograde flow of material at the surface of a polarized cell propels the cell forward \cite{Liu2015,Skupinski1980,Wilson2010,Maiuri2015,Bergert2015,Reversat2020}. This mechanism requires mechanical coupling of the cell surface with the environment, which can be generically expected to result in an effective friction of specific or unspecific origin \cite{Liu2015,Maiuri2015,Bergert2015,sens-PNAS,Reversat2020}. Retrograde flow is accompanied by anterograde transport to sustain recirculation of cellular materials, including structural and regulatory molecules \cite{Wilson2018}, see Fig.~\ref{fig1}(a). Interestingly, an analogous mode of locomotion develops collectively in sheets of cells during supracellular migration, which also involves planar polarization and coherent cellular flows characterized by pairs of vortices that organize retrograde and anterograde flows \cite{Shellard2019a,Shellard2020}, see Fig.~\ref{fig1}(b). Such a  mechanism has been observed in a variety of systems ranging from small cell clusters involving tens of cells \cite{Shellard2021,Shellard2018,Cetera2018} to developing embryos exhibiting large-scale morphogenetic collective motions of thousands of cells \cite{Saadaoui2020,Lawton2013,Rozbicki2015}, and from cohesive epithelial layers \cite{Cetera2018} to loosely associated cell collectives \cite{Shellard2021,Shellard2018}. 

The physical basis of fountain-flow migration for single cells has been extensively studied \cite{doi:10.1091/mbc.e04-10-0860,Bergert2015,Hawkins2011,Hawkins2009}. In this case, the forward anterograde flow takes place within the cytosol in the form of disassembled  cytoskeletal components or endocytic vesicles, and hence does not physically interact with the external environment. In cell clusters or tissues, however, the forward transport of cells in the sheet also interacts with the environment and mechanically opposes migration propelled by retrograde flow. Therefore, it is unclear how this supra-cellular dynamical organization with conflicting flows achieves efficient migration, as in neural crest clusters migrating over long distances and with cluster velocities comparable to individual cell velocities \cite{Shellard2019}. 

To understand supracellular fountain flow migration, we develop a continuum model that views the cell cluster as continuum viscoelastic and active 2D fluid in a time-dependent domain interacting frictionally with a substrate. Since we focus on the physical aspects of locomotion, we assume planar polarization of the cell sheet from the outset, which we refer to as cluster polarization to distinguish it from cell polarization. Inspired by neural crest cells (NCC) exhibiting  polarized contractility of a supracellular actomyosin cable at the periphery of the cell cluster  \cite{Shellard2018}, we introduce a gradient of line tension driving supracellular flow and  migration. An analogous mechanism operates during gastrulation of avian embryos \cite{Saadaoui2020}. Our model shows that the naive assumption of uniform and linear friction is incompatible with locomotion. To reconcile this fact with observations, we then consider biologically-motivated physical scenarios enabling competent fountain-flow supracellular migration.  

\section{Results}

\subsection{A first observation for a minimal crawler}
\label{first_obs}

\begin{figure}
	\centering
	\includegraphics[width=.85\linewidth]{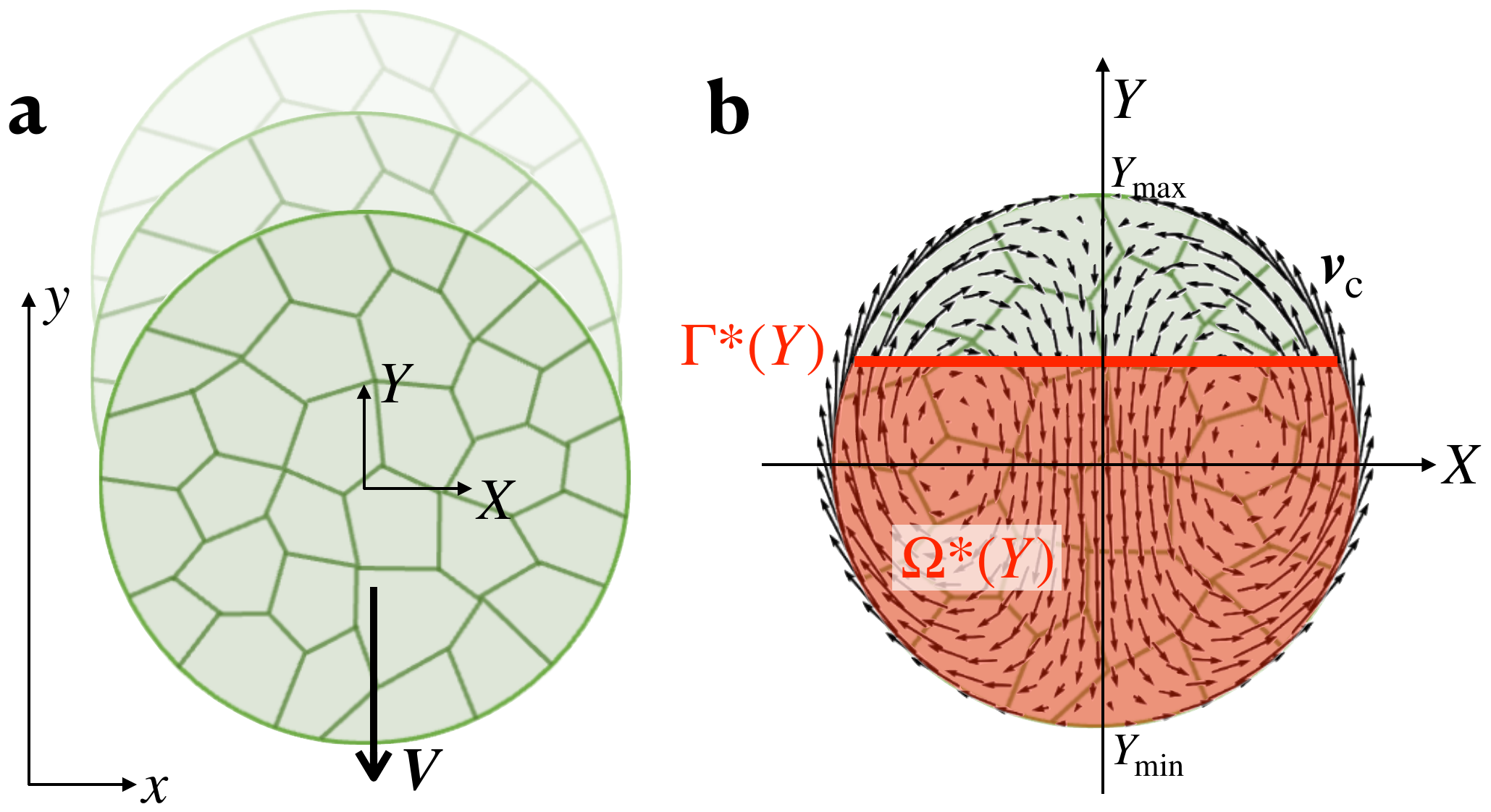}
	\caption{{\bf Analysis of a minimal crawler.} (a) Cellular cluster moving in the lab frame with velocity $\bm{V}$. (b) Velocity field $\bm{v}_{\rm c}$ in a frame moving with the crawler, along with geometric definitions used to conclude that $\bm{V}= 0$ when $\nabla\cdot\bm{v}_{\rm c}= 0$.}
	\label{fig11}
\end{figure}

To appreciate the fundamental challenge of fountain-flow 2D crawling, we first consider a minimal model for a fountain-flow crawler at steady-state, occupying a region $\Omega$ in the $(x,y)$ plane moving with velocity $\bm{V} = (0, V^y)$, which without loss of generality is along the $y$ axis. We view the cell assembly as a 2D incompressible fluid, a common assumption for flowing tissues, and denote by $\bm{v}(x,y)$ the velocity field within the crawler in the fixed frame of the substrate. We ignore for now how this internal flow  is generated. This field is divergence-free, $\nabla\cdot \bm{v} = 0$, and can be expressed in terms of the velocity field in a frame moving with the crawler, $\bm{v}_{\rm c}$, as  $\bm{v}(x,y) = \bm{v}_{\rm c}(X,Y) + \bm{V}$, Fig.~\ref{fig11}(a). 

The sliding of the tissue with respect to the substrate generates a frictional traction, which in the simplest case of a linearly viscous cell-substrate interface, takes the form $\bm{f}_{\rm fr} = -\eta\bm{v}$ with $\eta$ a constant friction coefficient. The total force acting on the crawler being zero, we have 
\begin{equation}
0 = \int_\Omega \bm{f}_{\rm fr} \,dS = -\eta \left(\int_\Omega \bm{v}_{\rm c} dS +  \vert \Omega \vert \bm{V}\right),
\end{equation}
where $\vert \Omega \vert$ is the area of the crawler. Using Fubini's theorem to compute this integral and focusing on the $y$ component, we can express the crawler velocity as
\begin{equation}
 V^y = - \frac{1}{\vert \Omega\vert}\int_{Y_{\rm min}}^{Y_{\rm max}} \left(\int_{\Gamma^*(Y)}  {v}_{\rm c}^y dX \right)dY,
 \label{Vy}
\end{equation}
where the interval $\Gamma^*(Y)$ is illustrated in Fig.~\ref{fig11}(b) and the  super-index denotes the velocity component along $y$. To compute the inner integral, consider now the subdomain $\Omega^*(Y)\subset \Omega$ bounded by $\Gamma^*(Y)$ and the lower part of the crawler's boundary $\partial\Omega$, Fig.~\ref{fig11}(b). Because of incompressibility and the divergence theorem, we have
\begin{equation}
0 = \int_{\Omega^*(Y)} \nabla\cdot\bm{v}_{\rm c}  \,dS = \int_{\partial \Omega^*(Y)} \bm{v}_{\rm c}\cdot \bm{n}\, ds = \int_{\Gamma^*(Y)} {v}_{\rm c}^y dX,
\label{compen}
\end{equation} 
where $\bm{n}$ is the outer unit normal of the boundary. To obtain the last expression, we have used that, because the shape of the steady-state crawler is constant in the moving frame, $\bm{v}_{\rm c}\cdot \bm{n} = 0$ over $\partial \Omega(Y)$, and also that the unit normal over $\Gamma^*(Y)$ is $\bm{n}= (0,1)$. Equation (\ref{compen}) reflects the notion that anterograde and retrograde net flows across a control line compensate. Combining this expression with Eq.~(\ref{Vy}), we conclude that the crawler velocity is zero irrespective of the internal cellular flow. 

This simple calculation clearly illustrates the fundamental difficulty of fountain-flow 2D crawling, which results from the conflict between retrograde and anterograde flows. By reconsidering its assumptions, this derivation also sheds light into alternative physical scenarios compatible with efficient fountain flow migration. Key assumptions leading to $\bm{V}= 0$ are (1) 2D incompressibility of the flow, (2) linear and uniform frictional cell-substrate interaction, and (3) no cellular self-propulsion. To lift these assumptions and explain the generation of a fountain flow from a polarized line tension, we develop next a more general theoretical model. Another assumption leading to $\bm{V}= 0$ not examined here is that the system is in a steady-state, which may not be the case in NCC \cite{Shellard2018} as a result of pulsating contractions. 

\subsection{Theoretical model of fountain flow migration}

We model the migrating tissue as a 2D compressible fluid characterized by a cell number density field $\rho$ and a velocity field $\bm{v}$ in the fixed frame of the substrate, which occupies a time-dependent domain $\Omega(t)$ with boundary $\partial \Omega(t)$. Unless needed, we drop in our notation the time dependence of the domain. Assuming for simplicity that cells do not divide or die, and hence their number is constant, local balance of cell number is expressed as $\partial_t\rho + \nabla\cdot(\rho\bm{v}) = 0$. We assume that cell rearrangements relax deviatoric elastic stresses and result in viscous 2D stresses of the form $\bm{\sigma}^\text{v} = 2\mu \nabla^\text{s} \bm{v}$, where $\mu$ is the viscosity and $\nabla^\text{s} \bm{v}$ is the symmetrized velocity gradient \cite{popovic2017active}. Such rearrangements do not relax isotropic elastic stresses, which we assume to depend on cell density as $\bm{\sigma}^\text{e} = g(\rho) \bm{I}$, where $g(\rho)$ is an equation of state for the tissue and $\bm{I}$ is the identity matrix. We consider $g(\rho) = k \left[1-(\rho/\rho_0)^{3/2}\right]$, where $k$ is a bulk modulus and $\rho_0$ is the equilibrium density of the fluid. This form of the equation of state is motivated by the physics of  epithelial monolayers \cite{Latorre2018}, but similar relations can emerge from the interplay between repulsion mediated by contact inhibition of locomotion (CIL) and co-attraction in loose embryonic cell associations such as NCC \cite{Shellard2019}, and the essential features of our results are insensitive to the form of $g(\rho)$, see Appendix \ref{EoS}. 

Given the low Reynolds number during tissue migration, balance of linear momentum in the tissue can be written as 
\begin{equation}
	- \nabla\cdot\bm{\sigma} = \bm{f}_{\rm fr}(\bm{v}) \;\;\;\;\;\;\mbox{in} \;\Omega, 
	\label{lin_mom}
\end{equation}
where $\bm{\sigma} = \bm{\sigma}^\text{e} + \bm{\sigma}^\text{v}$ is the total 2D stress, and $\bm{f}_{\rm fr}(\bm{v})$ is the velocity-dependent frictional traction with the substrate, see Fig.~\ref{fig1}(c).

The boundary is subjected to a non-uniform active line tension $\gamma(s)$ driving locomotion \cite{Shellard2019}, where $s$ is an arc-length parameter of $\partial \Omega$, see Fig.~\ref{fig1}(d). Such line tension generates a boundary traction, so that force balance at cluster edge is expressed by the boundary condition (Appendix \ref{Eq5})
\begin{equation}
	\bm{\sigma}\cdot\bm{n} = \partial_s (\gamma \bm{t}) =  (\partial_s \gamma)\,  \bm{t} - \gamma \kappa \, \bm{n} \;\;\;\;\;\;\mbox{on} \;\partial\Omega,
	\label{bnd}
\end{equation}
where $\bm{t}$ is the unit tangent vector to  $\partial\Omega$ along $s$, $\bm{n}$ the unit outward normal, and $\kappa$ is the curvature (positive for a convex part of the boundary). The last equality follows from the Frenet equations.  Finally, the evolution of the crawler domain is given by the relation $V_{\partial\Omega}(x,y,t) = \bm{v}(x,y,t)\cdot \bm{n}(x,y,t)$ for $(x,y)\in \partial\Omega(t)$, where  $V_{\partial\Omega}$ is the normal velocity of the boundary.

The total force acting on the crawler can be computed by integrating the frictional and boundary tractions 
\begin{equation}
	\bm{F} = \int_\Omega \bm{f}_{\rm fr} \,dS + \int_{\partial\Omega} \bm{\sigma}\cdot\bm{n} \,ds.
	\label{global_force}
\end{equation}
Because the crawler is self-propelled, this total force should vanish, as easily shown by replacing Eq.~(\ref{lin_mom}) in the first integral and invoking the divergence theorem. Furthermore, because the second integral is performed over a closed curve, Eq.~(\ref{bnd}) shows that it is zero, and hence each integral in Eq.~(\ref{global_force}) vanishes independently.

\subsection{Fountain flow locomotion driven by a gradient of line tension}

If $\gamma$ is uniform, Eq.~(\ref{bnd}) reduces to Laplace's law and the steady state of the system is a quiescent circle of radius $R^*$ and uniform density $\rho^*$, see Appendix \ref{R_star}. In contrast, if line tension is not uniform, Eq.~(\ref{bnd}) shows that  boundary tractions have tangential and normal components, both of which are cluster-polarized owing to the spatial variation of $\gamma(s)$. Besides perturbing the circular shape, these tractions generically lead to internal persistent flows as argued next. Because the elastic component of the stress is hydrostatic, the tangential component $(\partial_s \gamma)\,  \bm{t}$ can only be balanced by the viscous traction $\bm{\sigma}^\text{v}\cdot\bm{n}$, which in turn requires a sustained flow field with gradients at the edge. For instance, around point E in Fig.~\ref{fig1}(d), the velocity for a crawler polarized along $y$ is predominantly along $y$ and hence, according to Eq.~(\ref{bnd}),  $\mu \partial v^y_c/\partial x \approx \partial_s \gamma$. Because of symmetry, the velocity gradient has opposite sign at point W. 

To solve the flow generation and locomotion problem, i.e.~to solve for $\bm{v}(\bm{x},t)$, $\rho(\bm{x},t)$ and $\Omega(t)$ given initial conditions, material parameters ($k$, $\rho_0$, $\mu$, $\eta$) and the active line tension distribution $\gamma(s)$, we develop a computational method based on finite elements and an arbitrarily Lagrangian-Eulerian formulation to account for shape changes of the cluster \cite{donea2017arbitrary}. To introduce cluster polarization, we consider for simplicity  a linear profile of active line tension $\gamma(s) = \gamma_0 [1+  f(s)]$ where $\gamma_0$ is the average line tension and $f(s)$ the spatial variation. We consider $f(s) = 2\alpha (y(s)-Y_{\rm c})/L_y$ where $\alpha$ is a dimensionless measure of the tension gradient, $Y_{\rm c}=\int_\Omega y dS/\int_\Omega dS$ is the geometric center of the cluster  and $L_y$ is the length of the tissue in the $y$ direction.  We note that $\gamma(s)$ remains everywhere positive whenever $\vert\alpha\vert <1$. See \cite{Theveneau2010,Shellard2018,Shellard2021} for biological mechanisms of cluster polarization in NCC.

\begin{figure*}
	\includegraphics[width=.99\linewidth]{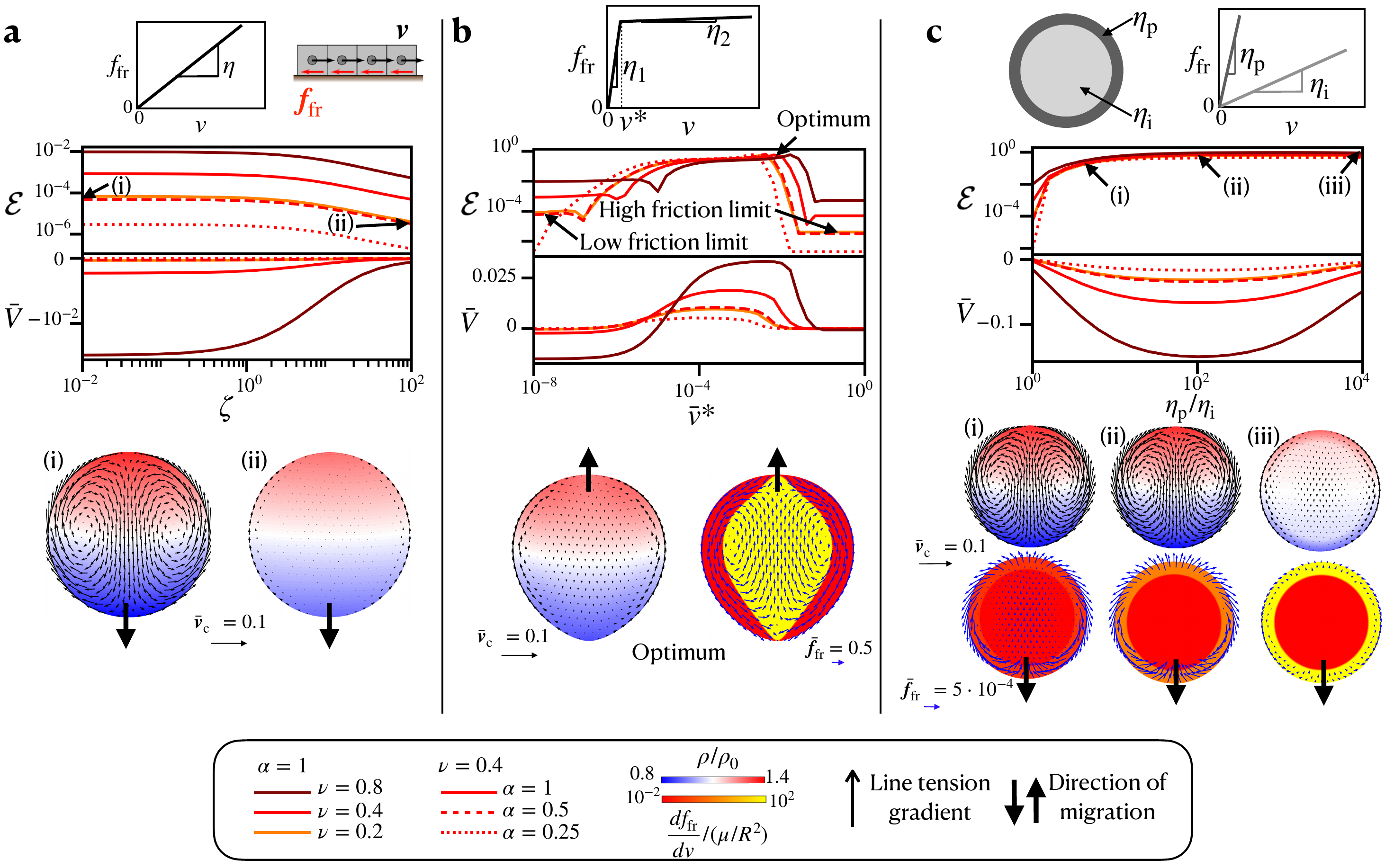}
	\caption{{\bf Effect of friction on fountain flow migration.} (a) Effect of dimensionless friction  $\zeta= R^2 \eta/\mu$. Top: Sketch of the linear traction-velocity relation. Middle: Lighthill efficiency $\mathcal{E}$ and signed dimensionless cluster speed $\bar{V}$ (positive along tension gradient) as a function of $\zeta$ for different dimensionless average line tensions $\nu$ (different colors) and line tension gradients $\alpha$ (continuous, dashed or dotted line). Here and elsewhere, overline velocities and tractions are dimensionless, normalized by $Rk/\mu$ and  $k/R$, respectively. Bottom: Illustration of fountain flow pattern at low (i) and high (ii) $\zeta$, where the arrows show the velocity field in the cluster frame $\bm{v}_{\rm c}$ and the colormap the normalized cell density field with white corresponding to the uniform density  $\rho^*$ of a cluster with constant line tension ($\alpha = 0$). Because cluster velocity is very small here, $\bm{v}_{\rm c}\approx \bm{v}$ and the negative of these velocity fields is proportional to the traction exerted by the substrate on the cell cluster. (b) Effect of nonlinear rheology of the cell-substrate interface. Top: Sketch of the nonlinear friction-velocity relation characterized by two slopes and a threshold velocity $v^*$. Middle: Lighthill efficiency $\mathcal{E}$ and signed cluster velocity $\bar{V}$ as a function of dimensionless threshold $\bar{v}^*$. Bottom: Velocity field in the cluster frame $\bm{v}_{\rm c}$ and density field (left) and self-organized frictional pattern and traction field $\bm{f}_{\rm fr}$ (right) at optimal efficiency. (c) Effect of spatially patterned friction. Top: Sketch of the heterogeneous pattern of friction. Middle: Lighthill efficiency $\mathcal{E}$ and signed cluster velocity $\bar{V}$ as a function of friction contrast. Bottom: Velocity field in the cluster frame $\bm{v}_{\rm c}$ and density field (top) and frictional pattern and traction field $\bm{f}_{\rm fr}$ (bottom) for representative friction contrasts. In all cases, $\mathcal{E}$ is represented in logarithmic scale whereas $\bar{V}$ in linear scale. All representative velocity and friction maps correspond to $\nu = 0.2$ and $\alpha = 1$.}
	\label{fig2}
\end{figure*}

We first consider the simplest case of uniform linear friction, $\bm{f}_{\rm fr}(\bm{v}) = -\eta\bm{v}$, initialize the system at the steady-state for $\alpha=0$, suddenly apply a tensional gradient $\alpha>0$, and follow the dynamics. After a transient, the system reaches a steady state with a sustained fountain flow and a net motion in the $y$ direction Movie 1. In agreement with observations on NCC \cite{Shellard2019}, the direction of migration is opposite to the line tension gradient, and hence is driven by the retrograde flow at the periphery of the tissue and dragged by the internal anterograde stream,  Fig.~\ref{fig2}(a).

\subsection{Role of friction magnitude, rheology and heterogeneity}

To examine the physical aspects of this migration mode, we varied the dimensionless numbers  $\zeta = R^2 \eta/\mu = (R/\ell_h)^2$, where $\ell_h = \sqrt{\mu/\eta}$ is the hydrodynamic length, $\nu = \gamma_0 /(kR)$, and $\alpha$. These numbers control the competition between viscosity and friction, between line tension and density elasticity, and the strength of the gradient, respectively. For the sake of discussion, we view cluster size $R$, elastic modulus $k$ and viscosity $\mu$ as fixed, and vary friction or the properties of the contractile cable. 

To quantify migration efficiency, we adapt Lighthill's efficiency to the present context  to define
\begin{equation}
	\mathcal{E}[\bm{v}] = \bm{V}\cdot \int_{\Omega} \bm{f}_{\rm fr}(\bm{V})  dS \left/ 
	\int_{\Omega} \bm{v}\cdot\bm{f}_{\rm fr}(\bm{v}) dS\right. .
\end{equation}
This quantity compares the power required to drag a passive crawler at the actual migration velocity $\bm{V}$ with the power performed by the active crawler on the external medium \cite{Friedrich2018}, here a frictional substrate. To further interpret this quantity, for a linear frictional law it takes the form $\mathcal{E}[\bm{v}] = |\bm{V}|^2\int_{\Omega} \eta  dS \left/ 
\int_{\Omega}  \eta |\bm{v}|^2 dS\right.$, and hence its square root is the  ratio between the migration velocity and a weighted average of the velocity of the particles within the cluster.

For $\zeta\ll 1$, the flow pattern exhibits system-size coherent flows with two prominent vortices. Instead, for $\zeta\gg 1$, the retrograde flow is restricted to a thin peripheral region, separated by an internal layer from the anterograde flow occupying most of the interior of the cluster, see Fig.~\ref{fig2}(a). This can be understood by noting that large $\zeta$ corresponds to small hydrodynamic length and hence strong screening by friction of stresses in the cell sheet. Furthermore, the magnitude of dimensionless cell velocities $\bar{\bm{v}}_c = \bm{v}_c/(Rk/\mu)$ decreases with increasing $\zeta$, see Fig.~\ref{fig2}(a). The system also develops density gradients, which are more pronounced for small relative friction $\zeta$. 

We then examine efficiency by changing dimensionless friction $\zeta$ by four orders of magnitude and considering different properties of the contractile cable given by $\nu$ and $\alpha$, see Fig.~\ref{fig2}(a). Different parameter combinations lead to noticeably different flow patterns, Movie 2. In all cases, efficiency plateaus for $\zeta<1$, and then rapidly reduces in the high-friction regime following a power law with exponent $\sim -0.8$. Efficiency is remarkably sensitive to moderate changes in $\nu$ and $\alpha$, increasing by more than two orders of magnitude when either of these parameters increases by four-fold. For migrating clusters of NCC, we estimate $\nu \simeq 0.1$ to 0.8, $\alpha \simeq 0.5$, and $\zeta \simeq 0.025$, see Appendix \ref{params}, leading to $\mathcal{E}\simeq 10^{-7}$ to $10^{-4}$ and $\bar{V} = V/(Rk/\mu)\simeq 10^{-6}$ to $10^{-3}$, which assuming $R_0 = 50\;\upmu$m, $k = 2$ nN/$\upmu$m and $2\mu = 10^2$ nN min/$\upmu$m, corresponds to a cluster velocity of $2\cdot10^{-6}$ to $2\cdot10^{-3}$ $\upmu$m/min, orders of magnitude smaller than the observed velocities of about 2 $\upmu$m/min \cite{Shellard2018}. Furthermore, $\mathcal{E}\sim 10^{-4}$ means that the cluster velocity is about 100 times smaller than internal particle velocities in the cluster frame, at odds with the comparable velocity of NCC cells and the cluster \cite{Shellard2018}. This result is not surprising since the density variations within the cluster remain moderate, and hence the system is not very far from the incompressible limit, which as discussed in Section \ref{first_obs} exhibits no motion. Furthermore, if we slightly modify the model by making the frictional tractions proportional to density, $\bm{f}_{\rm fr}(\bm{v}) = -\eta \rho \bm{v}$, then analogous arguments to those in Section \ref{first_obs} can be made to conclude that $V=0$ in the compressible case. Taken together, these results show that linear uniform friction is incompatible with robust and significant fountain-flow supracellular locomotion.

To reconcile our theoretical predictions with observations, we examine whether a more complex rheology of the cell-substrate interaction  can lead to more efficient migration. Indeed, the collective effect of ensembles of transient molecular bonds under force can lead to nonlinear rate-dependent frictional interactions \cite{Srinivasan2009,sens-PNAS}. Nonlinear Bingham-type frictional interactions have been shown to enable more efficient crawling \cite{DeSimone2013}. We consider that $\bm{f}_{\rm fr}$ is still antiparallel to $\bm{v}$, but now the magnitude $f_{\rm fr}(v)$ is a bilinear relation, saturating the frictional force beyond a threshold $v^*$, Fig.~\ref{fig2}(b) and Appendix \ref{bilinear}. We set the two slopes such that initially we are in a high-friction regime,  $\zeta_1 = R^2 \eta_1/\mu = 10^2$, and after the threshold the incremental friction coefficient corresponds to the low-friction limit,  $\zeta_2 = R^2 \eta_2/\mu = 10^{-2}$. We then vary the dimensionless threshold $\bar{v}^* = v^* /(Rk/\mu)$ and track efficiency and cluster velocity for different combinations of parameters characterizing the contractile peripheral cable, $\nu$ and $\alpha$, Fig.~\ref{fig2}(b) and Movie 3. 

When $v^*$ is very low or very high, we expectedly recover the low- and high-friction regimes of Fig.~\ref{fig2}(a). For intermediate $v^*$, the system develops heterogeneous distributions of tangent friction coefficient $\eta_{\rm t}(v) = df_{\rm fr}/dv$, which enable efficient migration with $\mathcal{E}$ approaching 1. At the optimum, the system self-organizes with low $\eta_{\rm t}(v)$ in the periphery and higher $\eta_{\rm t}(v)$ in the center. In this regime of intermediate $v^*$, both efficiency and cluster velocity are maximal but the direction of migration reverses and is along tension gradients. This effect can be rationalized noting that the heterogeneous friction modifies the mechanical balance between opposing flows in such a way that now the retrograde flow is the central stream whereas anterograde flow is along the cluster edges. Still, it is somewhat surprising in that the system moves in the direction of increasing tension. Remarkably, changes in the  active tension distribution, i.e.~changes in $\nu$ and $\alpha$, have a strong effect on the efficiency in the low/high-friction limits, but very little effect in the optimal efficiency.

Thus,  nonlinear frictional  leads to self-organization into spatially segregated regions of saturated friction at the periphery of the cell cluster, which makes fountain flow locomotion efficient and robust with respect to the details of the line tension gradient. However, the emerging behavior contradicts NCC supracellular migration in that its direction is along the tensional gradient. Furthermore, by a mechanism of CIL, adhesive interactions between  NCC and the substrate have been shown to localize at the periphery of cell clusters  \cite{Roycroft2018,Shellard2019,Shellard2020}. Therefore, CIL changes the behavior of peripheral cells, whose frictional interaction with the substrate is expected to be much larger than that of internal cells, opposite to the self-organized frictional pattern resulting from nonlinear friction, Fig.~\ref{fig2}(b). 

To examine the effect spatial patterning of friction as a result of CIL, we then consider a linear but heterogeneous frictional law $\bm{f}_{\rm fr}(\bm{v}) = -\eta(\bm{x}) \bm{v}$, where the peripheral region has friction $\eta_{\rm p}$ and the inner region $\eta_{\rm i}$ with $\eta_{\rm p}>\eta_{\rm i}$, Fig.~\ref{fig2}(c). To study the effect of friction contrast, we choose internal friction  to be in the low-frictional limit of Fig.~\ref{fig2}(a) such that $\zeta_{\rm i} = \eta_{\rm i} R^2/\mu= 10^{-2}$ and vary $\zeta_{\rm p}$ between this value and  $10^{2}$. In all cases, the cluster moves opposite to the tension gradient, in agreement with NCC. When  $\eta_{\rm p}/\eta_{\rm i}=1$, our results recover the large sensitivity to the pattern of active tension $\gamma(s)$ and low efficiency of Fig.~\ref{fig2}(a). However, as the friction contrast increases, $\mathcal{E}$ approaches 1 and becomes mildly dependent on the form of active tension. The flow, density and traction patterns are also modified, with tractions becoming increasingly peripheral and cell velocities smaller, Fig.~\ref{fig2}(c) and Movie 4. Although efficiency plateaus as $\eta_{\rm p}/\eta_{\rm i}$ increases and is rather insensitive to $\nu$ and $\alpha$, the non-dimensional cluster velocity $\bar{V}= V/(Rk/\mu)$ attains a maximum when $\eta_{\rm p}/\eta_{\rm i} \approx 100$ and significantly increases with baseline tension and its gradient. Thus, whereas increased uniform friction is always detrimental to migration, see Fig.~\ref{fig2}(a), increased peripheral friction strongly enhances migration efficiency and robustness, and can lead to fast migration for intermediate friction contrasts and strongly contracting and polarized peripheral cables. With the parameters used above for NCC and $\eta_{\rm p}/\eta_{\rm i} \approx 100$, we obtain $\mathcal{E}\simeq 0.4$ to $0.6$ and ${V}\simeq 2\cdot 10^{-2}$ to $2\cdot 10^{-1}$ $\upmu$m/min, only ten times smaller than measured cluster velocities. For  $\eta_{\rm p}/\eta_{\rm i} \approx 10$ the results are comparable.

\begin{figure*}
	\includegraphics[width=.85\linewidth]{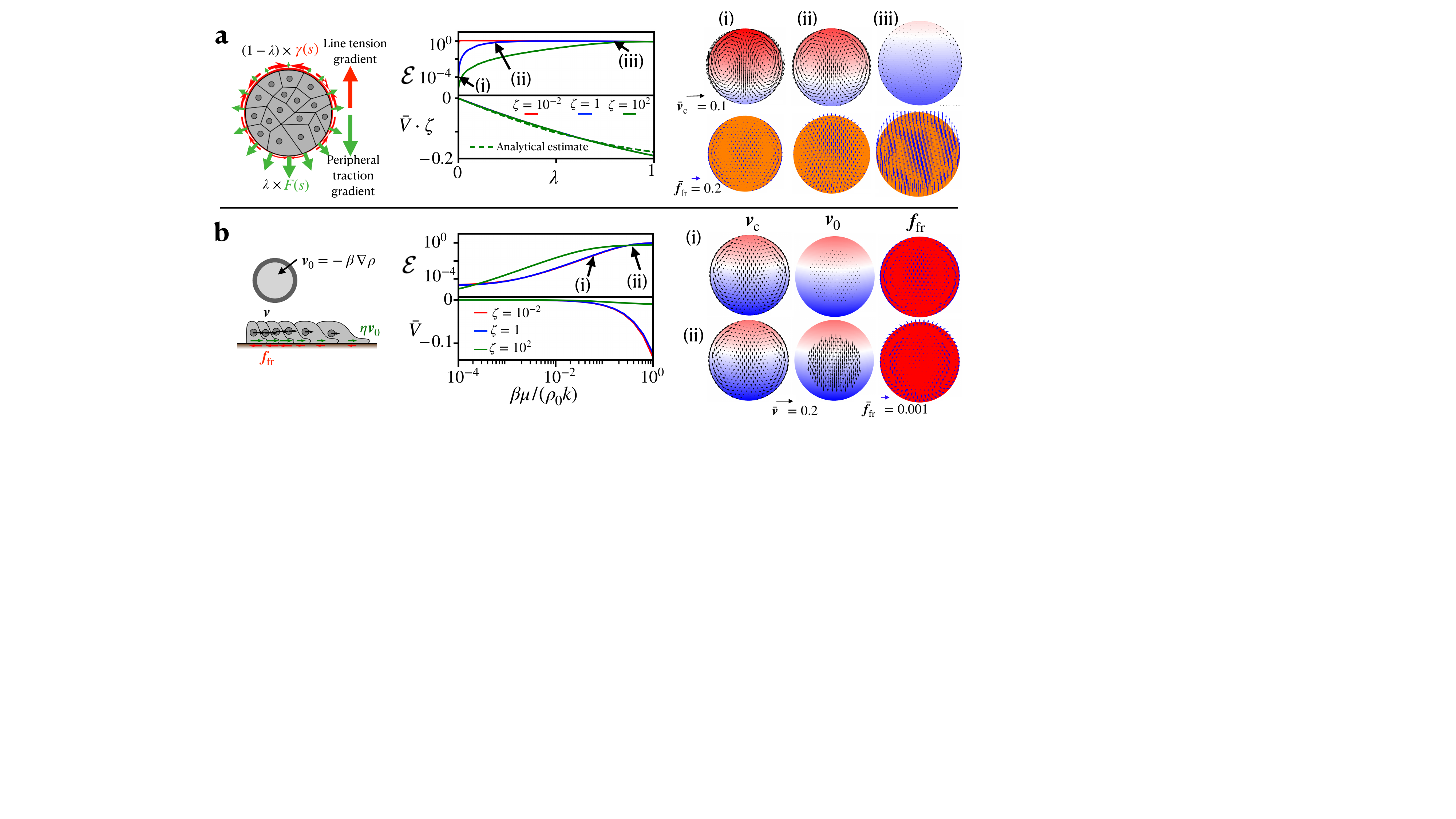}
	\caption{{\bf Effect of polarized crawling activity on fountain flow migration.} (a) Left: Alternative model accounting for peripheral crawling forces pointing along the outward normal (green arrows). This mechanism can also be cluster-polarized and co-exist with  peripheral line tension with opposite polarity (red arrows). Middle: Lighthill efficiency $\mathcal{E}$ and signed normalized cluster velocity $\bar{V}$ multiplied by $\zeta$ as a function of the relative weight $\lambda$ between the line tension and the peripheral traction gradients for three dimensionless friction coefficients $\zeta$ (velocity curves overlap). The dashed line is the analytical estimate discussed in the text. Right: Velocity field in the cluster frame $\bm{v}_{\rm c}$ and density field (top) and passive traction field $\bm{f}_{\rm fr}$ (bottom) for representative values of $\lambda$.
		(b) Left: Alternative model accounting for polarized crawling of interior cells following a minimal Toner-Tu model \cite{Toner1995} with coupling parameter $\beta$ relating self-propulsion velocity $\bm{v}_0$ and density gradient. Middle:  Lighthill efficiency $\mathcal{E}$ and signed normalized cluster velocity $\bar{V}$ as a function of dimensionless coupling $\beta$  for three dimensionless friction coefficients $\zeta$. Right: Velocity field in the cluster frame $\bm{v}_{\rm c}$ and density field (left), self-propulsion velocity field $\bm{v}_0$ (middle) and traction field $\bm{f}_{\rm fr}$ (right) for representative values of $\beta$. In all cases, $\nu = 0.2$ and $\alpha = 1$.}
	\label{fig3}
\end{figure*}

\subsection{Role of individual cell migration}

Besides a gradient in supracellular contractility, it has been suggested that NCC migration may also depend on individual cell polarization and self-propulsion \cite{Shellard2020,Shellard2019a}. According to CIL, peripheral crawling tractions induced by polymerization in lamellipodia at the edge of the cluster point outwards as shown in Fig.~\ref{fig3}(a) (green arrows). These peripheral tractions can also be cluster-polarized, and result in the following boundary condition 
\begin{equation}
	\bm{\sigma}\cdot\bm{n} = F(s) \, \bm{n} \;\;\;\;\;\;\mbox{on} \;\partial\Omega.
	\label{bnd2}
\end{equation}
The mechanics of this alternative polarization mechanism are very different. First, Eq.~(\ref{global_force}) also holds but in general both integrals are different from zero. Second, Eq.~(\ref{bnd2}) can be rewritten as $g(\rho)\, \bm{n} + \bm{\sigma}^{\rm v}\cdot\bm{n} =  F(s) \, \bm{n}$. Consequently, it is possible to balance a gradient of $F(s)$ solely with a gradient of hydrostatic elastic stress, and hence of cell density, without shear dissipation caused by velocity gradients. Thus, in principle, we could set the velocity field in the cell frame to zero, $\bm{v}_{\rm c}= 0$, and find the cluster velocity from Eq.~(\ref{global_force}). For a linear and uniform friction-velocity relation, this leads to the explicit formula 
\begin{equation}
	V = \frac{1}{\eta A}\int_{\partial\Omega} F(s) n^y\,ds,
	\label{V_trac}
\end{equation}
where $n^y$ is the $y$ component of the unit outward normal and $A$ is the surface area of the cluster. Considering the same kind of spatial distribution for $F(s)$ as we did for $\gamma(s)$ characterized by a mean traction $F_0$ and a dimensionless gradient $\alpha$, we obtain the explicit formula $\bar{V} = -\alpha F_0/(\zeta k)$ depending on the new dimensionless number $F_0/k$, or $V=-\alpha F_0/(\eta R)$, see Appendix \ref{formulavF}. Our simulations agree remarkably well with this rationale, and we find that a gradient of peripheral traction $F(s)$ produces a nearly rigid body motion given by the previous formula with no internal cellular flows, and hence no fountain-flow recirculation, and uniform drag inside the cluster, see Fig.~\ref{fig3}(a,iii).

Interestingly, due to the antagonism between regulators of the actin cytoskeleton Rac1 and RhoA \cite{Guilluy:2011aa}, the two peripheral active mechanisms given by Eqs.~(\ref{bnd}) and (\ref{bnd2}) can be expected to exhibit opposite polarity, see Fig.~\ref{fig3}(a), and hence act in concert. To study this possibility, we considered the same line tension distribution as earlier, $\gamma(s) = \gamma_0 \left[1+f(s)\right]$, and a related peripheral normal traction distribution  $F(s) = F_0 \left[1-f(s)\right]$. We then studied a combination of these two active peripheral mechanisms. To ease the comparison, we chose $F_0 = \gamma_0 /R_0$ and considered an interpolation between both boundary conditions by scaling these effects as $(1-\lambda) \gamma(s)$ and $\lambda F(s)$ for $0\le\lambda\le 1$. In the limit of small shape changes (not assumed in our simulations), $\kappa \approx 1/R_0$ and hence
\begin{equation}
	\bm{\sigma}\cdot\bm{n}  \approx (1-\lambda) \gamma_0 f'(s)\, \bm{t} - \frac{\gamma_0}{R_0} f(s) \bm{n} + \frac{\gamma_0}{R_0}(2\lambda-1)\, \bm{n},
\end{equation}
where the last terms only contribute to a uniform compression or extension of the cluster. Thus, the main effect of changing $\lambda$ is tuning the strength of the tangential part of the boundary condition. We found that with increasing contribution of the peripheral traction (increasing $\lambda$), the fountain flow progressively disappears, and efficiency rapidly increases reaching values close to 1, with high friction delaying the plateau, see Fig.~\ref{fig3}(a) and Movie 5. Furthermore, the cluster velocities are large and closely follow $\bar{V} = -\lambda \alpha F_0/(\zeta k)$ over the entire range of $\lambda$ as predicted in the limit of pure peripheral traction with intensity $\lambda F_0$. Thus, these results show that for uniform and linear friction (1) fountain flow results from polarized peripheral line tension and (2)  cluster velocity is overwhelmingly dominated by the contribution from peripheral and polarized traction. Quantitatively, we find that for $\alpha \simeq 0.5$, $\nu \simeq 0.4$, $\zeta \simeq 0.025$ and the previous estimates $R_0 = 50\;\upmu$m, $k = 2$ nN/$\upmu$m and $2\mu = 10^2$ nN min/$\upmu$m, we achieve the observed velocity of 2 $\upmu$m/min for $F_0 \approx 0.1$ nN/$\upmu$m. Noting that $F_0 \approx 3$ nN/$\upmu$m in expanding cell monolayers \cite{blanch2017effective} (see also  Appendix \ref{params}), this figure of the peripheral traction is reasonable.

In addition to individual cell polarization at the periphery of the cluster, CIL also plays a role within the cell cluster by polarizing internal cells, which may develop cryptic lamellipodia, from denser regions towards emptier regions \cite{Shellard2020}. Such a mechanism can be introduced in our model by considering a self-propulsion velocity of cells $\bm{v}_0$, or equivalently an active crawling traction $\eta \bm{v}_0$, so that the frictional force can be expressed as $\bm{f}_{\rm fr} = -\eta(\bm{v}-\bm{v}_0)$. For simplicity, we consider $\eta$ constant. To determine $\bm{v}_0$, we consider a minimal version of the Toner-Tu theory of flocking \cite{Toner1995}, such that $\bm{v}_0 = -\beta \nabla \rho$ in interior cells, see Fig.~\ref{fig3}(b). We find that, as the coupling coefficient $\beta$ becomes larger, cells become increasingly polarized along the negative of the tensional gradient, hence cancelling the friction of the anterograde flow opposing migration, see Movie 6, and hence increasing migration efficiency to values close to 1 as well as velocity. Thus, individual cell polarization inside of the cluster following a negative cell density gradient can also contribute to efficient fountain flow supracellular migration.

\section{Summary and conclusions}

In summary, we have shown that supracellular fountain-flow  migration, prevalent during development, is the result of gradients of peripheral line tension (as produced by a cluster-polarized supracellular actomyosin cable) but not to gradients of peripheral normal traction (as produced by cluster-polarized protrusive activity). Supracellular fountain-flow migration is more difficult than its cellular counterpart because anterograde flows resist motion. As a result, if the mechanical interaction with the substrate is uniform and linear friction, migration velocity and efficiency are extremely  small. Fast, efficient and robust migration can be rescued by a nonlinear frictional interaction, by larger peripheral friction as suggested by increased adhesions at the edges of NCC clusters, or by single cell polarization and self-propulsion as dictated by CIL. Thus, supracellular migration may resort to any of these complementary mechanisms or a combination of them to robustly achieve migration in changing environments. These physical rules can also be applied to optimize locomotion in engineered active matter \cite{Ross:2019tu}.

\section*{Acknowledgements}

The authors acknowledge the support of the European Research Council (CoG-681434), and the Ministry of Science, Innovation and Universities of Spain (project PID2022-142178NB-I00 funded by MICIU/AEI/ 10.13039/501100011033 and ERDF/EU). MA acknowledges the Generalitat de Catalunya (ICREA Academia prize for excellence in research). IBEC and CIMNE are recipients of a Severo Ochoa Award of Excellence. 

\appendix

\section{Sensitivity to the choice of equation of state} \label{compareelastcoeff}
\label{EoS}

\begin{figure}[h!]
	\includegraphics[width=.7\linewidth]{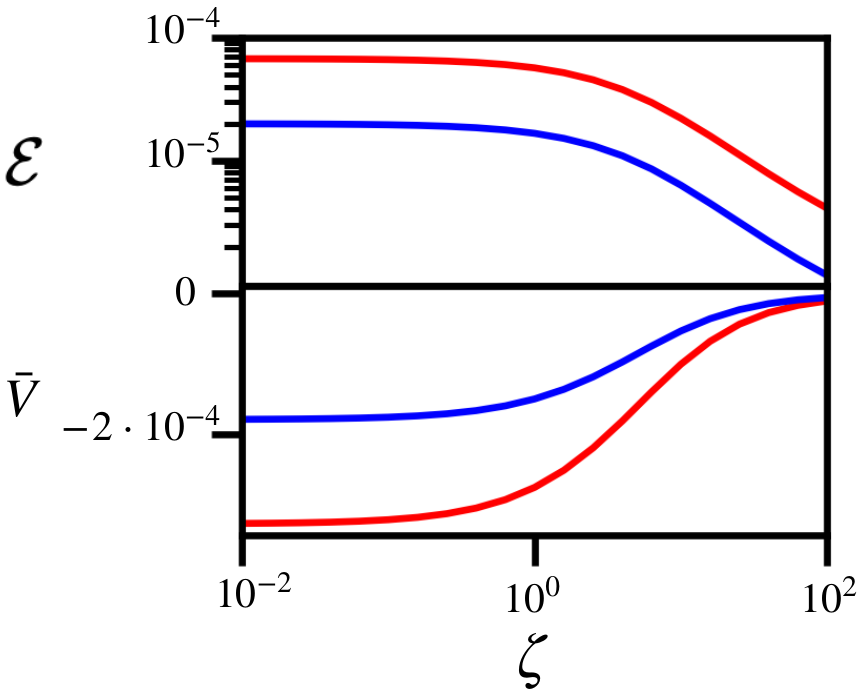}
	\caption{Comparison between alternative forms of the equation of state. The red line represents the efficiency and velocity using $g(\rho) = k \left[1-(\rho/\rho_0)^{3/2}\right]$, whereas the blue line is for $g_2(\rho) = k \left[1-(\rho/\rho_0)^{2}\right]$. In all simulations, $\nu = 0.2$ and $\alpha = 1$.}
	\label{supfig1}
\end{figure}

The elastic stress is isotropic, and takes the form $\bm{\sigma}^\text{e} = g(\rho) \bm{I}$. In our simulations, we consider $g(\rho) = k \left[1-(\rho/\rho_0)^{3/2}\right]$ motivated by the apparent elasticity of epithelial monolayers \citep{Latorre2018}. However,  a quadratic expression of the kind $g_2(\rho) = k \left[1-(\rho/\rho_0)^{2}\right]$ is more common. To show that the specific form of the equation of state of the tissue  does not alter the qualitative result of our simulations, we compare two sets of simulations  using $g(\rho)$ and $g_2(\rho)$. Fig.~\ref{supfig1} shows that, even if the simulations provide quantitatively different results, there are no qualitative differences, and hence our main conclusions are unaffected by the choice of equation of state.

\section{Derivation of Eq.~(\ref{bnd})} 
\label{Eq5}

The power input associated to line tension can be computed by integrating line tension times the rate of elongation of the boundary
\begin{equation} \label{appcont}
	\mathcal{P}_C[\gamma] = \int_{\partial\Omega} \gamma \cdot \dot{ds}.
\end{equation}
Let $\bm{x}(u)$ be a parametrization of the boundary curve $\partial\Omega$, so that $\bm{\tau} = \partial_u \bm{x}$ is the natural tangent vector to the boundary edge. We thus have $ds = | \bm{\tau} | du$. Time-differentiating this expression, we obtain \begin{equation}  \label{apptau}
\dot{ds} = \frac{\bm{\tau} \cdot \dot{\bm{\tau}}}{| \bm{\tau} |} du, 
\end{equation}
where changing the order of differentiation, the time derivative of the tangent vector can be written as $\dot{\bm{\tau}} = \partial_t \partial_u \bm{x} = \partial_u \partial_t \bm{x} = \partial_u \bm{v}$. We can thus rewrite Eq.~\eqref{appcont} as
\begin{equation}
	\mathcal{P}_C[\gamma, \bm{v}] = \int_{\partial \Omega} \gamma \,\bm{t} \, \partial_u  \bm{v}\,du,
\end{equation}
where $\bm{t} = (1/|\bm{\tau}|)\bm{\tau}$ is a tangent unit vector. Using the product rule, the power functional can be rewritten as
\begin{align} \label{appgammatau}
	\mathcal{P}_C[\gamma, \bm{v}] = &
 \int_{\partial\Omega} \partial_u \left( \gamma \bm{t} \cdot \bm{v} \right) \, du  \nonumber \\
	& - \int_{\partial\Omega} \partial_u \left( \gamma \bm{t} \right) \cdot \bm{v} \, du.
\end{align}
The first integral  is zero since $\partial\Omega$ is a closed curve. Using the product rule and the Frenet equation on the second integral, we have
\begin{align}
	\mathcal{P}_C[\gamma, \bm{v}] 
	& =  -\int_{\partial \Omega} \left( \partial_u \gamma  \bm{t}  + \gamma \partial_u \bm{t} \right) \cdot \bm{v} \, du \nonumber \\
	& = -\int_{\partial \Omega} \left( \partial_s \gamma  \, \bm{t}  - \gamma \kappa \, \bm{n}\right) \cdot \bm{v} \, ds,
\end{align}
where $\kappa$ is the curvature of the boundary $\partial \Omega$ and ${\bm{n}}$ is the unit outer normal. Because the stress power at the boundary takes the form 
\begin{align}
\int_{\partial \Omega} \bm{v}\cdot \bm{\sigma} \cdot \bm{n} \,ds,
\end{align}
we deduce Eq.~(\ref{bnd}) from the principle of virtual power. 

\section{Equilibrium radius $R^*$ under uniform line tension} \label{apprstar}
\label{R_star}

When line tension is uniform $\gamma$, the equilibrium state is a circular domain characterized by the equilibrium radius $R^*$ and a constant density.  Invoking the boundary condition at $\partial \Omega$ along the normal direction and disregarding out-of-equilibrium contributions to the stress, we find that 
\begin{equation}
	k \left(1 - \frac{R_0^3}{R^3} \right) = -\frac{\gamma}{R}.
\end{equation}
The equilibrium radius is then the solution of the polynomial equation
\begin{equation} \label{eqr}
	R_0^3 - (R^*)^3 = \frac{\gamma}{k}(R^*)^2.
\end{equation}

\section{Parameters from experiments} \label{expparam3}
\label{params}

In this appendix, we provide reasonable estimates for the model parameters according to the literature. The radius $R$ of the cell cluster is taken from \cite{Shellard2018} and is equal to $R$ = 50 $\upmu$m. From  \cite{Shellard2018, scarpa2015cadherin}, we obtain the typical velocity of the cluster of about $V \sim 2 \;\upmu$m/min. For the elastic bulk  modulus $k$, we consider $k = 2$ mN/m $=2$ nN/$\upmu$m from \cite{Latorre2018}. Reference \cite{blanch2017effective} provides estimates for several parameters such as the viscous coefficient $2\mu^{\rm bulk} = 10^5 - 10^6$ Pa min for an epithelial monolayer 7 microns tall, and hence $2\mu \sim 10^3$ nN min/$\upmu$m. For a loose collective such as ours, we consider a 10-fold smaller viscosity, $2\mu \sim 10^2$ nN min/$\upmu$m. Taking a hydrodynamic length of 300 $\upmu$m, we obtain a friction coefficient of $\eta \sim 10^{-3}$ nN min/$\upmu$m$^3$. 

To estimate line tension $\gamma_0$, we use observations from \cite{Shellard2018} of cluster compression under pulsations of line tension. According to this reference, a  $R_0=50$ $\upmu$m cluster reduces its radius to approximately $R=40$ $\upmu$m upon tension increase. If we assume that this new radius is the equilibrium radius $R^*$, then  Eq.~\eqref{eqr} provides an estimate of line tension as $\gamma = k [R_0^3 - (R^*)^3]/(R^*)^2 \sim$ 80 nN. This figure is not far from the estimate $\gamma_0\sim 10$ nN according to \cite{nier2015tissue}. 

We estimate the peripheral traction forces from data of migrating cell monolayers with peripheral active traction and roughly straight edges. In this quasi-1D setting, the peripheral tractions $F_0$ of our model correspond to the integral along the expanding direction of the measured cell-substrate traction near the edge, which closely follow the tension in the tissue. According to \cite{blanch2017effective}, for expanding monolayers  $F_0 \sim 3$  nN/$\upmu$m. This figure, representative of a highly polarized leading edge, should be an upper bound. Using these parameters, we estimate the dimensionless numbers  $\zeta \sim 0.025$, $\nu \sim 0.1$ to $1$.

\section{Implementation of bilinear friction relation} \label{appbifric}
\label{bilinear}

To model nonlinear friction, we consider traction-velocity relations of the form $\bm{f}_{\rm fr}(\bm{v}) = -  {f}_{\rm fr}({v})  \bm{v}/v$ where ${f}_{\rm fr}({v})$ is the modulus of the frictional traction and $v=|\bm{v}|$. For such a nonlinear relation, we define the tangent friction coefficient as
\begin{equation} 
	\eta({v}) = \frac{d {f}_{\rm fr}}{dv}({v}).
\end{equation}
To model a  stick-slip behavior, we consider a situation in which cells strongly interact with the substrate via cell-matrix adhesions at small sliding velocities  (stick), but such that  when the sliding velocity is larger than a threshold adhesions weaken, allowing  cells to slide  without increasing significantly the cell-substrate tangential traction (slip). We define ${f}_{\rm fr}({v})$ as a bilinear function given by an initial stick friction $\eta_1$ and a slip friction $\eta_2 \ll \eta_1$ past a critical sliding velocity $v^*$:
\begin{equation} \label{stickslip}
	\eta({v}) = \begin{cases}
		\eta_{1} &\quad\text{if } v < v^*,\\
		\eta_2 &\quad\text{if } v > v^*.
	\end{cases}
\end{equation}
To regularize the corner between these two regimes, we consider a smoothened version of Eq.~\eqref{stickslip} given by
\begin{equation} \label{stickslipcon}
	\eta({v}) = \begin{cases}
		\eta_{1} &\,\text{if } v \leq v^{**},\\
		\eta_2 + \frac{q}{v} &\,\text{if } v > v^*,\\
		av^3 + bv^2 + cv +d + \frac{p}{v} &\,\text{otherwise},\\
	\end{cases}
\end{equation}
where $v^{**} = 0.9 v^*$, $q$ is the minimum frictional force to transition to the slip regime and $a$, $b$, $c$, $d$ and $p$ are parameters chosen to ensure continuity and differentiability of the function.

\section{Explicit formula for a circular cluster velocity} \label{formulavF}

Given the general formula for the cluster velocity under peripheral traction forces $F$ in Eq.~(\ref{V_trac}), we derive next an explicit expression when the domain $\Omega$ is a circle of radius $R$. We remind that $F(s) = F_0[1-f(s)]$, where $f(s) = 2\alpha (y(s) -Y)/L_y$, with $y(s)$  the vertical position, $Y$  the position along $y$ of the center of the cluster and $L_y$  the vertical length. For a circular domain with radius $R$, we have that $f(s) = \alpha \sin(s/R)$. Then, given that $n^y = \sin (s/R)$ and $A = \pi R^2$, we can explicitly compute the integral on the circle to obtain
\begin{equation} \label{v}
	V = -\frac{\alpha F_0}{\eta R},
\end{equation}
or in non-dimensional terms (considering as characteristic speed $Rk/\mu$)\begin{equation}
	\bar{V} = - \alpha \frac{\bar{F}_0}{\zeta},
\end{equation}
where $\bar{F}_0 = F_0/k$ is a new dimensionless parameter.

\section*{Movie captions} \label{movies3}

\bigskip	

\noindent{\bf Movie~1:} \quad Evolution of a simulation for homogenous and uniform friction with parameters $\nu = 0.4$, $\alpha = 1$ and $\zeta = 0.01$. Simulations are performed until the system reaches a steady state, where the velocity respect to the cluster (black arrows) is tangential at the boundary and density (color map) is concentrated at the cluster rear (at top, in red). Even if it is difficult to appreciate, the entire cluster is migrating downwards.

\bigskip

\noindent{\bf Movie~2:}  \quad Steady states of clusters with  uniform linear friction for parameters $\nu = 0.4$ and $\alpha = 1$ and varying dimensionless friction coefficient $\zeta$.  (Left) Black arrows represent the velocity field in the frame of the cluster and color map shows density. (Right) Blue arrows represent frictional forces and the color map shows $\zeta$. The plot indicates the efficiency and dimensionless velocity as a function of  $\zeta$.

\bigskip	

\noindent{\bf Movie~3:} \quad	Steady states of clusters with  stick-slip friction for parameters $\nu = 0.4$, $\alpha = 1$, $\zeta_1 = 100.0$ (stick), $\zeta_2 = 0.01$ (slip) and varying dimensionless critical slip velocity $v^*$.  (Left) Black arrows represent the velocity field in the frame of the cluster and color map shows density. (Right) Blue arrows represent frictional forces and the color map shows the dimensionless tangential friction coefficient. The plot indicates the efficiency and dimensionless velocity as a function of  $v^*$.

\bigskip	

\noindent{\bf Movie~4:} \quad		Steady states of clusters with heterogeneous linear friction, larger in the periphery, for parameters $\nu = 0.4$, $\alpha = 1$, $\zeta_i = 0.01$ (internal region) and varying ratio between internal and peripheral friction $\eta_p/\eta_i$.  (Left) Black arrows represent the velocity field in the frame of the cluster and color map shows density. (Right) Blue arrows represent frictional forces and the color map shows the dimensionless  friction coefficient. The plot indicates the efficiency and dimensionless velocity as a function of $\eta_p/\eta_i$.
\bigskip	

\noindent{\bf Movie~5:} \quad	Steady states of clusters with uniform linear friction and a combination of a peripheral polarized line tension $(1-\lambda)\gamma(s)$ and oppositely polarized outward peripheral tractions $\lambda F(s)$, with parameter $\lambda$ weighing their relative contribution. Parameters are $\nu = 0.4$, $F_0 = 0.4$, $\alpha = 1$, $\zeta = 1$ and varying $\lambda$.  (Left) Black arrows represent the velocity field in the frame of the cluster and color map shows density. (Right) Blue arrows represent passive frictional forces. The plot indicates the efficiency and dimensionless velocity as a function of $\lambda$.

\bigskip	

\noindent{\bf Movie~6:} \quad		Steady states of clusters with uniform linear friction and a combination of a peripheral polarized line tension and self propulsion of internal cells according to a minimal Toner-Tu model where self-propulsion velocity $\bm{v}_0$ is proportional to the negative of the  gradient of density with coupling coefficient $\beta$. Parameters are $\nu = 0.4$,  $\alpha = 1$, $\zeta = 1$ and varying $\beta$.  (Left) Black arrows represent the velocity field in the frame of the cluster and color map shows density. (Middle) Black arrows represent the self-propulsion velocity field $\bm{v}_0$. (Right) Blue arrows represent frictional forces. The plot indicates the efficiency and dimensionless velocity as a function of $\beta$.

\bibliographystyle{apsrev4-1}

\end{document}